\documentclass[11pt]{article}
\usepackage{hyperref} 
\usepackage{geometry}
\usepackage[utf8]{inputenc}
\usepackage{xcolor}
\usepackage{graphicx}
\usepackage{amsmath}

\newenvironment{affiliations}{%
    \setcounter{enumi}{1}%
    \setlength{\parindent}{0in}%
    \slshape\sloppy%
    \begin{list}{\upshape$^{\arabic{enumi}}$}{%
        \usecounter{enumi}%
        \setlength{\leftmargin}{0in}%
        \setlength{\topsep}{0in}%
        \setlength{\labelsep}{0in}%
        \setlength{\labelwidth}{0in}%
        \setlength{\listparindent}{0in}%
        \setlength{\itemsep}{0ex}%
        \setlength{\parsep}{0in}%
        }
    }{\end{list}\par\vspace{12pt}}

\renewenvironment{abstract}{%
    \setlength{\parindent}{0in}%
    \setlength{\parskip}{0in}%
    \bfseries%
    }{\par\vspace{0pt}}
\def\trap{\text{TRAPPIST-1}}
\def\wat{\text{H$_2$O}}

\newcommand{\beginsupplement}{%
        \setcounter{table}{0}
        \renewcommand{\thetable}{S\arabic{table}}%
        \setcounter{figure}{0}
    \renewcommand{\figurename}{Supplementary Figure}
     }

\title{Inward Migration of the TRAPPIST-1 Planets as Inferred From Their Water-Rich Compositions}

\author{Cayman T. Unterborn$^{1,3}$, Steven J. Desch$^{1}$, Natalie R. Hinkel$^{2}$, Alejandro Lorenzo$^{1}$}

\begin{document}

\maketitle
\begin{affiliations}
 \item School of Earth and Space Exploration, Arizona State University, Tempe, AZ 85287, USA
 \item Department of Physics \& Astronomy, Vanderbilt University, Nashville, TN 37235, USA
 \item cayman.unterborn@asu.edu
\end{affiliations}
%%%%%%%%%%%%%%%%%%%%%%%%%%%%%%%%%%%%%%%%%%%%%%%%%%%%%%%%%%%%%

\begin{abstract}
Multiple planet systems provide an ideal laboratory for probing exoplanet composition, formation history and potential habitability. For the TRAPPIST-1 planets, the planetary radii are well established from transits \cite{Gill16,Gill17}, with reasonable mass estimates coming from transit timing variations \cite{Gill17,Wang17} and dynamical modeling \cite{Quar17}. The low bulk densities of the TRAPPIST-1 planets demand significant volatile content. Here we show using mass-radius-composition models, that \trap{}f and g likely contain substantial ($\geq50$ wt\%) water/ice, with b and c being significantly drier ($\leq15$ wt\%). We propose this gradient of water mass fractions implies planets f and g formed outside the primordial snow line whereas b and c formed inside. We find that compared to planets in our solar system that also formed within the snow line, TRAPPIST-1b and c contain hundreds more oceans worth of water. We demonstrate the extent and timescale of migration in the \trap{} system depends on how rapidly the planets formed and the relative location of the primordial snow line. This work provides a framework for understanding the differences between the protoplanetary disks of our solar system versus M dwarfs. Our results provide key insights into the volatile budgets, timescales of planet formation, and migration history of likely the most common planetary host in the Galaxy.
\end{abstract}

The derivation of a planetary composition from only its mass and radius is a notoriously difficult exercise because of the many degeneracies that exist. The geophysical and geochemical behavior of a planet is extremely sensitive to such factors as the size of the iron core, the mantle mineralogy, and the location of phase boundaries within any rock and ice layers \cite{Dorn15,Unte16}. For astrobiological applications it is crucial to constrain the exact amount of surficial water a planet contains. Yet current models assume only pure iron cores and an Earth-like composition for the mantles, and often assume {\it either} an iron core plus silicate mantle {\it or} a silicate planet plus ice mantle \cite{Zeng16}. While useful for broadly constraining rocky versus volatile-rich composition, current mass-radius constraints often fail to meaningfully quantify the specific planetary composition \cite{Dorn15,Unte16}. For example, \cite{Quar17} constrained the masses of the \trap{} planets using dynamical stability arguments, and found they were compatible with compositions between 0\% and 100\% water ice. The mass-radius fitting of \cite{Gill17} and \cite{Wang17} provided similar, but still uncertain, constraints. We argue that simultaneous mass-radius-composition fitting of all the TRAPPIST-1 planets, using the context from the planetary system as a whole, allows better quantification of the allowable structures and mineralogies given the mass-radius measurements and their uncertainties. We, therefore, analyzed the interior structures and mineralogies of the the six best-constrained \trap{} planets (b-g) using the ExoPlex code (Methods). ExoPlex computes the mass of planet given the input radius with the assumed composition. The identity of the less-dense component of each of the \trap{} planets is almost certainly either ${\rm H}_{2}{\rm O}$ liquid/ice (because of its cosmochemical abundance) or a gaseous envelope \cite{Gill17}. On large planets with radius greater than $1.6 \, R_{\oplus}$ radii, extended H$_2$ atmospheres significantly lower the bulk density of the planet \cite{Weis14,Roge15}, but we consider this unlikely for any of the \trap{} planets (Methods). Ruling out an atmosphere entirely is nearly impossible, but we proceed under the assumption that the \trap{} planets all lack extended atmospheres and that the transit radii measure their solid surfaces and the volatile component of each planet is liquid water and/or water-ice. For each modeled planet, we adopt a bulk Fe/Mg and Si/Mg ratio, ${\rm H}_{2}{\rm O}$ mass fraction and total radius (see Methods for details). 
\begin{figure}
    \begin{center} \includegraphics[width=12cm]{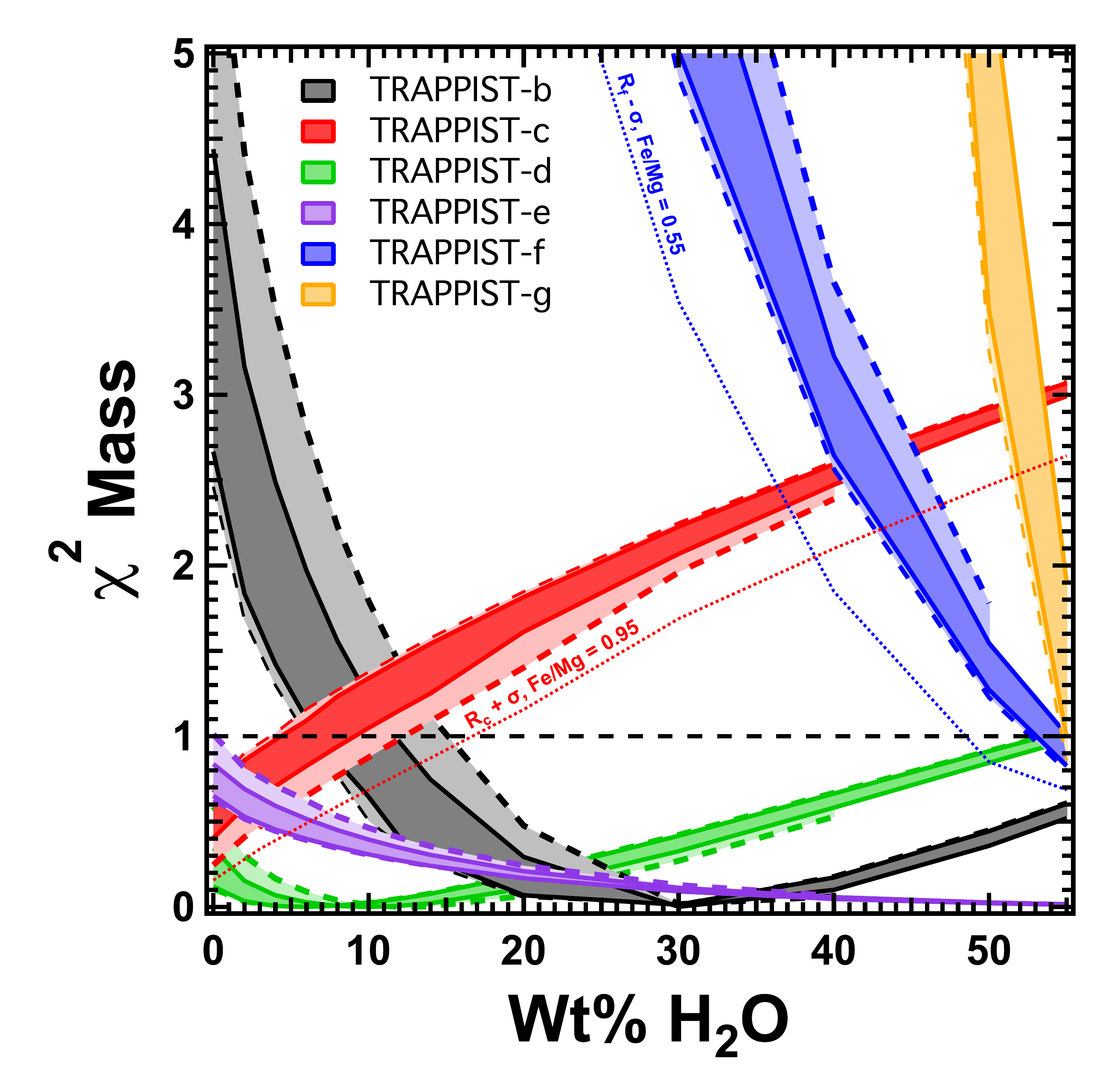}
    \end{center}
    \caption{Modeled $\chi^{2}$ goodness-of-fit for the \trap{} planets as a function of the relative \wat{} fraction in weight percent added to the system. Lighter bands represent those models with 0.5 $<$ Fe/Mg $<$ 1.3 (a proxy for the relative size of a planet's core and mantle) while the darker bands are those models with 0.55 $<$ Fe/Mg $<$ 0.9 (Thick dashed lines represent Fe/Mg = 1.3). This subsample are those stars with compositions within 1$\sigma$ of the average Fe/Mg for 930 FGK stars from the Hypatia Catalog\cite{Hink14} (see Methods) within the reported metallicity range of \trap{} range -0.04 $\leq$ [Fe/H] $\leq$ 0.12 \cite{Gill16}. For those modeled planets with Fe/Mg = 1.3, no solution for core-mantle-ice shell size was found to suffice compositional and water mass fraction constraints. These areas are left blank.}\label{fig:chi}
\end{figure}
We explored the phase space of potential \trap{} composition by calculating $\chi^2$ in mass for each planet (Figure 1) as a function of water mass fractions of $f_{\rm H2O}$ from 0 - 55 wt\% (note: the value is $< 0.1$wt\% for Earth \cite{Mott07}) and across an Fe/Mg range characteristic of stars of similar metallicity (Figure 1; Methods and Supplementary Figures 2 and 3). A planet's Fe/Mg is roughly a measure of the relative ratio of core to mantle \cite{Dorn15,Unte16,Unte17}. For each planet, the modeled mass, $M_{\rm mod}$, is compared to the observed mass, $M_{\rm obs}$, and a goodness-of-fit parameter: $\chi^2 \equiv (M_{\rm mod} - M_{\rm obs})^2 / \sigma_{\rm M}^2$, is found. Here, $\sigma_{\rm M}$ is the observational uncertainty in the mass. Uncertainties in radius are also considered by increasing or decreasing the input total radius of the modeled planet accordingly. We adopt the $M_{\rm obs}$ and $\sigma_{\rm M}$ data of \cite{Wang17} for this study. These masses are consistent with the dynamical orbital stability studies \cite{Quar17}. 

To fit \trap{}b at the $1\sigma$ level requires $7\leq f_{\rm H2O}\leq$ 12 wt\%. In contrast, \trap{}c is fit at the $1\sigma$ level by $f_{\rm H2O}$ {\it less than} 10 wt\% for the same Fe/Mg range. Planets d and e are equally well fit ($\chi^2 \geq 1$) by all considered compositions due to their high uncertainties in mass, however, $\chi^2$  for planet d begins to rise above 1 as $f_{\rm H2O}$ increases above 50 wt\%. Compared to the inner planets, \trap{}f and g are much more water-rich and are best fit only with water fractions greater than $\leq50$ wt\% water, regardless of Fe/Mg. If we assume that water fractions increase with orbital radius as observed in our solar system, we can use the composition of planet c to constrain that of planet b, such that both planets likely have between 7 and 10 wt\% water. Assuming \trap{}c and f are 1$\sigma$ larger and smaller in radius, respectively, only represents a 5 wt\% increase and decrease in best-fit water fraction, increasing our calculated upper-limit water fraction for \trap{}b and c to 15 wt\% \wat{}. Thus, within the current uncertainties, our model result that the \trap{} system contains relatively ``dry'' inner planets and wet outer planets is robust and is consistent with the findings of \cite{Gill17}. While the reported masses of the \trap{} planets may be revised with longer-time TTV data, our conclusion of a relatively ``dry'' inner and ``wet'' outer \trap{} system is consistent with the previous mass estimates of \cite{Gill17}. Furthermore, this result and line of reasoning is independent of how the \trap{} planets acquired their water, be it from formation beyond the snow line or due to late delivery via giant impacts \cite{Raym04}.
\begin{centering}
\begin{figure}
    \centering
    \includegraphics[width=.6\linewidth]{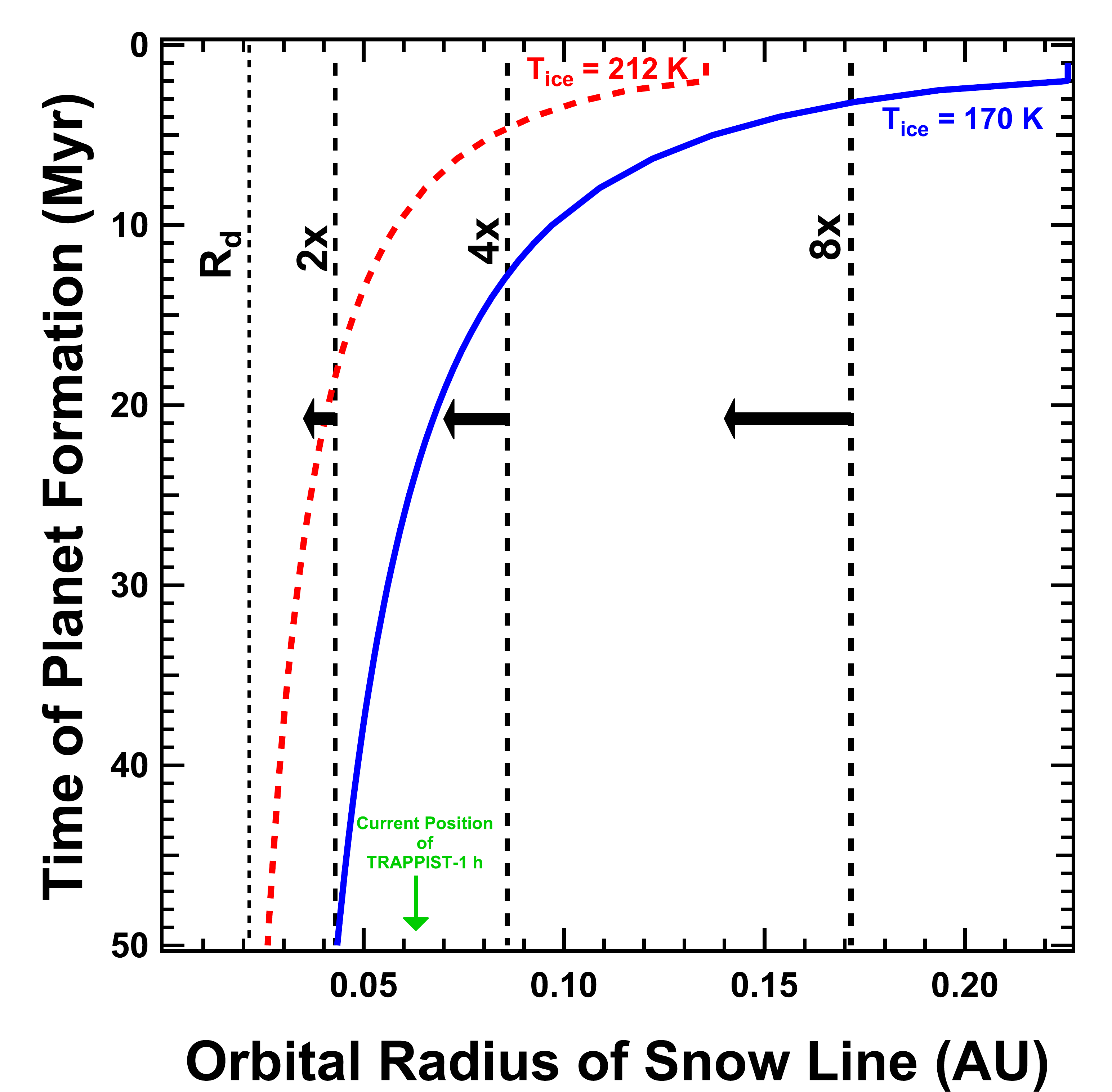}
    \caption{The orbital radius of our modeled water snow line as a function of time of planet formation. Blue lines represent the minimum pre-migration orbital radius of \trap{}d for various migration distance factors. If \trap{}d formed at $>$ 0.15 AU, it would have to have formed in $<$ 5 million years.}
    \label{fig:snowline}
\end{figure}
\end{centering}
All of the \trap{} planets currently orbit well inside the primordial snow line of their protoplanetary disk and thus are unlikely to have formed in-situ as in \cite{Quar17} (Methods). Indeed, radial migration of planets is commonly inferred in planetary systems \cite{Liss11,Fabr14,Stef15}, including the \trap{} system (Methods). From our modeled gradient of water fraction in the \trap{} disk, we infer then the innermost planets (b and c) formed inside a snow line and the outermost planets (f, g, and h) formed outside it and migrated to their current location (Methods). Due to the current uncertanties in the mass of planets d and e, our mass-radius-composition results are consistent with the primordial snow line being anywhere between the orbits of planets \trap{}c and f. For convenience and demonstration purposes, we proceed assuming the primordial snow line was present at the pre-migration orbital radius of planet d. We calculate the position of the snow line, the location in the disk where the temperature is such that ${\rm H}_{2}{\rm O}$ condenses as ice or remains as vapor. We calculate this temperature to be $T = 212$ K, which which is greater than that of the solar nebula (170 K) due to the greater surface density of M-dwarf disks compared to the solar nebula (Methods). We calculate the radial location of the snow line by assuming a passively heated, flared disk as modeled by \cite{Chia97}. We calculate location of $r_{snow}$ to be:
\begin{equation}
r_{\rm snow} \approx 0.06 \, \left( \frac{ M_{\star} }{ 0.08 \, M_{\odot} } \right)^{-1/3} \, 
                        \left( \frac{ L_{\star} }{ 0.01 \, L_{\odot} } \right)^{2/3} \, {\rm AU},
\label{eq:rsnow}
\end{equation}
where $M_{\star}$, $M_{\odot}$, $L_{\star}$, and $L_{\odot}$ are the mass and luminosity of the star and Sun, respectively (see Methods). While the stellar mass is fixed, the stellar luminosity of an M dwarf decreases steadily throughout the $\sim 10 - 30$ Myr typical lifetime of its protoplanetary disk \cite{Bara02}. We use the stellar evolution models of \cite{Bara02} to estimate the luminosity of \trap{} at various epochs. At age 10 Myr where $L_{\star} \approx 1 \times 10^{-2} \, L_{\odot}$, $r_{\rm snow} \approx 0.06 \, {\rm AU}$; however $r_{\rm snow}$ moves inward as $L_{\star}$ decreases over time (Figure 2). During the first 10 Myr of the evolution of the \trap{} protoplanetary disk, the snow line was outside the present orbital radius of all the planets, including h. Unless the disk persisted for substantially longer than 50 Myr, the snow line was always beyond the current orbital radius of \trap{}d. Inclusion of accretional heating in the disk would move the snow line even further out, requiring even greater migration than proposed here. The compositional arguments for migration reinforce strong evidence from the large masses and orbital resonances of the TRAPPIST-1 planets (Methods) that they underwent considerable inward migration. The degree of inward migration is uncertain because planet accretion timescales in M dwarf disks are unknown.

If the \trap{} planets formed within 10 Myr in their protoplanetary disk, from a mix of rock (inside the snow line) or rock and ice (outside the snow line), then their orbital radii must have decreased due to later migration by a factor $\sim3$ assuming the primordial snow line was present at the orbit of \trap{}d (Figure 2). In this scenario, if the planets formed more rapidly, within 3 Myr of the disk's lifetime, and if they reflect the composition of the disk at that time, then their orbital radii must have decreased by a factor of $\sim6$. Only if the planets took $\gg 20$ Myr to form could we infer that they migrated by less than a factor of two of their current orbits. Theoretical work\cite{Lee16} has suggested that that the \trap{} planets formed as late as $\sim$40 Myr if there is enough gas drag to prevent planetesimal growth until the gas begins to disperse. If planet formation took this long, and if TRAPPIST-1f formed just outside the snow line, it is possible to reconcile the planet compositions without any migration. We view this as highly unlikely, though, in light of the very rapid formation of planets in the solar system, our inference of faster planet growth in the \trap{} system (Figure 2), and the other arguments in favor of migration (Methods). At any rate, our modeling provides an outline for constraining the degree of migration as the \trap{} planet masses are better constrained, particularly those of planets d and e.
\begin{figure}
    \centering
    \includegraphics[width=0.6\linewidth]{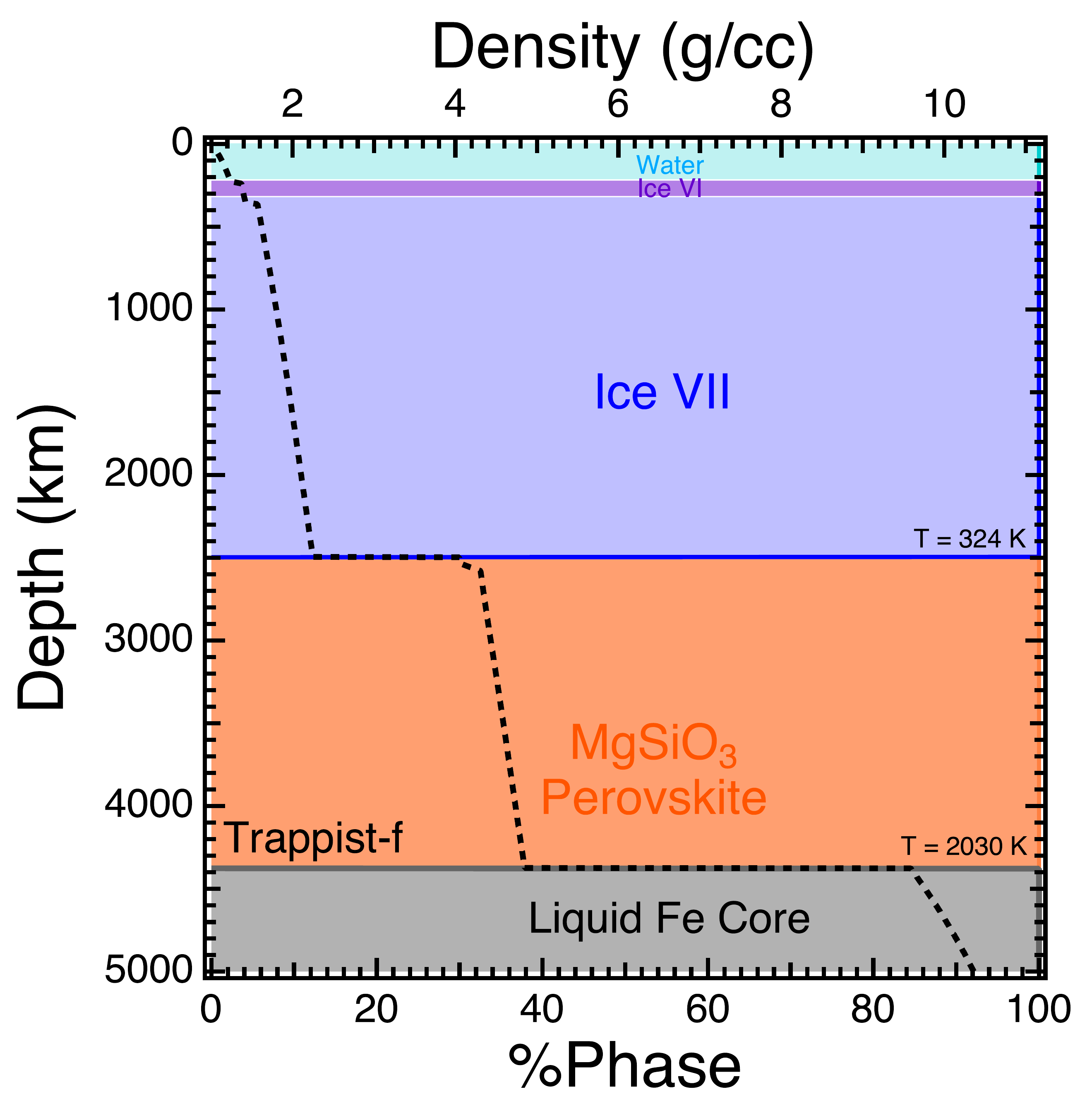}
    \caption{Phase diagram with depth as modeled with the ExoPlex mass-radius-composition calculator for the best-fit interiors of \trap{} f (Fe/Mg = 0.8, Mg/Si = 1 and 50 wt\% \wat{}; See Methods). Density as a function of depth is shown as a dashed line. Adiabatic temperatures at the bottom each major boundary layer are included for reference assuming thermal boundary layers at the water-mantle and core-mantle boundary (See Methods). Note that due to the increased pressure at the water-mantle boundary ($\sim$25 GPa), no upper mantle and transition zone mineral phases are present in \trap{}g. Due its the lower mass (and thus central pressure), no post-perovskite is present as well.}
    \label{fig:Mineralogy}
\end{figure}
Here we provide a theoretical framework of self-consistent mass-radius-composition calculations and astrophysical modeling for characterizing volatile-rich, multiple planet systems. Applying this framework to the \trap{} system, if planet d formed at or near the snow line, it must have migrated inward by a factor $\sim 2-4$, with the exact value depending on when the planets formed. Even this crude estimate of this value allows us to estimate of the surface density in \trap{}'s protoplanetary disk. We find it must have been similar to the ``minimum-mass solar nebula'' model of the average of M stars in the {\it Kepler} database \cite{Gaid17}, and much higher than in the solar nebula (Methods). As a consequence, planet formation and migration in all M dwarf disks is most likely more rapid than in our own Solar System's protoplanetary disk. The lower but still high value of $f_{\rm H2O}$ in planets b and/or c suggests that the snow line was not a sharp boundary, and/or that planets accreted substantial material from beyond the snow line after formation. Planets within M-dwarf systems, then, are likely to contain significantly more water than the Earth, or any planetary body in our solar system. Liquid water may exist below a high pressure ice layer\cite{Noac16} or tidal heating may cause temperatures to be above that of ice\cite{Barr18}. Even if planets are warm enough to contain no ice, given enough surface water the overbearing pressure will enough to suppress decompression mantle melting \cite{Kite09} and transition directly to a mineralogy where only lower-mantle minerals are stable below the water-mantle boundary (Figure 3). Given these factors, rocky planets with water fractions greater than Earth may not behave geochemically and geophysically similar manners as the Earth. Furthermore, continental crust is unlikely to reach above the water layer. With no exposed land, key geochemical cycles including the draw down of carbon and phosphorus into oceanic reservoirs from continental weathering will be muted, thus limiting the size of the biosphere. As such, while these planets may be habitable in the classical definition, any biosignature observed from these planets system may not be fully distinguishable from abiotic, purely geochemical sources. Thus, while M-dwarfs may be the most common habitable planet-host in our Galaxy, they may be the toughest on which to detect life.
\bibliography{main.bbl}

\newpage
\section{Methods}
\beginsupplement
\subsection{Retention of Atmospheres}
We consider the presence of significant atmospheres as unlikely for any of the \trap{} planets for a variety of reasons. First, even a small amount of H$_2$/He would have very noticeable effects. Namely, in an atmosphere of mass $\sim10^{-6} \, M_{\oplus}$ ($\sim1$ bar surface pressure for an ``Earth-like'' planet), the 1 microbar level, or the effective planetary photosphere\cite{Lope12} would lie at an altitude $\sim 4400$ km. An atmosphere such as this would significantly increase the radius of any of the \trap{} planets. If the \trap{} planets had accreted and retained extended atmospheres, at least some of them might be expected to have densities $<$1 g cm$^{-3}$, similar to Kepler 11f \cite{Bede17}. The H$_2$/He mass would have to be implausibly fine-tuned across the entire \trap{} system to lower the bulk densities to even a small extent. Second, the H$_2$/He atmospheres accreted by planets as small as the \trap{} planets are easily lost. The X-ray/ultraviolet (XUV) irradiation from \trap{} will cause $\sim10^{-4} \, M_{\oplus}$ of atmospheric mass loss on each planet. The proto-atmosphere accretion simulations of \cite{Stoe15} show that only planets $> 1 M_{\oplus}$ accrete atmospheres this massive. We find that planet c alone might have accreted an atmosphere more massive than might be lost by XUV irradiation; but c is also the densest planet and the one that shows the least evidence for retention of an atmosphere. Third, observations of other exoplanets strongly suggest a threshold radius $\approx 1.6 \, R_{\oplus}$ for significant reduction of inferred bulk density \cite{Weis14,Roge15}, and the \trap{} planets, all with radii $< 1.1 R_{\oplus}$, are well below the threshold. It also is likely all the \trap{} planets could  easily  replenish  water  vapor  atmospheres.  H$_2$O  atmospheres, however, will have a smaller scale height compared to H$_2$/He, and thus the presence of such atmospheres would not change the inferred radii substantially.

\subsubsection{H$_2$/He Atmosphere}
Whether any of the \trap{} planets can retain an extended H$_2$/He atmosphere that changes its inferred radius depends on how much atmosphere it can accrete from its disk, and how much atmosphere is lost over time by heating by XUV heating. 
We calculate the erosion of planetary atmospheres by XUV heating from the central star using the formalism of \cite{Lope12}. The energy-limited mass loss rate is
\begin{equation}
\dot{M} \approx \frac{ \epsilon \, \pi \, F_{\rm XUV} \, R_{\rm XUV}^3 }{ G M_{\rm p} \, K_{\rm tide} } \, ,
\end{equation}
where $F_{\rm XUV}$ is the flux of XUV photons, $R_{\rm XUV} \approx R_{\rm p}$ is the radius at which the atmosphere becomes optically thick to XUV photons \cite{Wats81,Erka07}, the efficiency with which XUV energy is converted into atmospheric loss is $\epsilon \approx 0.1$ \cite{Lope12}, and $K_{\rm tide} \approx 1$ is a small correction to account for the fact that gas must escape only to the Hill sphere to be lost. 
Other studies\cite{Whea17}, using a similar treatment, calculated the mass of atmosphere that could be lost if the mass loss was energy-limited and the efficiency by which absorbed XUV energy was converted into atmospheric escape was 10\%, using XMM-Newton measurements of the XUV fluxes from \trap{}. We update their calculations, scaling to an age of 8 Gyr \cite{Burg17}, and using the mass and radius for each planet as reported by \cite{Wang17}.  We find the following atmosphere mass losses over 8 Gyr for the \trap{} planets: 0.050, 0.013, 0.022, 0.021, 0.010, 0.005, and 0.011 Earth masses, for planets b through h, respectively.  In other words, atmospheres less massive than about 0.005-0.05 Earth masses are easily lost due to X-ray/UV irradiation over the age of the \trap{} system.

These are to be compared to the masses of hydrogen each planet might have accreted. According to the proto-atmosphere accretion models\cite{Stoe15}, planets $< 0.5 M_{\oplus}$ accrete $\ll 10^{-4} \, M_{\oplus}$ of gas. Only b, c and g need be considered. Without loss, planets b and g would today have an atmosphere $< 1 \times 10^{-5} \, M_{\oplus}$, so no extended atmosphere can be expected to remain on either planet. Planet c, however, might accrete an atmosphere $\sim 3 \times 10^{-4} \, M_{\oplus}$, marginally greater than the amount that can be lost. Because of its greater mass, planet c is the only planet in the \trap{} system that could retain a portion of an H$_2$/He proto-atmosphere accreted from its disk. But planet c is actually the densest of the \trap{} planets, and the one that shows the least evidence for retention of an H$_2$/He atmosphere. It seems unlikely that the \trap{} planets could retain H$_2$/He atmospheres. This is consistent with the finding that planets with radii $< 1.6$ R$_\oplus$ generally do not show evidence for inflation by H$_2$/He atmospheres. 

\subsubsection{H$_2$O Atmosphere}
It is likely all the \trap{} planets contain could easily replenish water vapor atmospheres. However,  H$_2$O atmospheres will have a smaller scale height compared to H$_2$/He atmospheres, and thus the presence of such atmospheres would not change the inferred radii substantially. Depending on the temperature, the maximum pressure at the base of an atmosphere can be estimated at 1-100 bar (at higher pressures the vapor condenses). The upper level of the atmosphere (from a transit perspective) is about 1 microbar \cite{Lope12}. Therefore, the maximum thickness of an H$_2$O atmosphere is 14-18 scale heights. Assuming $H = k T / (m g)$, imposing $T = 300$ K, we calculate $H$ = 31.5 km for \trap{}g, meaning that the radius of g is inflated by no more than 580 km, or 0.09 R$_\oplus$. Even with a maximal H$_2$O vapor atmosphere, g has a radius of at least 1.04 R$_\oplus$, and a density half of Earth's. Thus, planet g is still ice-rich.  Similar (but less precise) results are obtained for planet f.  If planets b and c possessed substantial H$_2$O atmospheres, they would likely have lower ice fractions and our results again represent an upper limit. Therefore we consider it a robust conclusion that even with H$_2$O atmospheres that b and c are relatively ice-poor and g (and f) are relatively ice-rich.

The ExoPlex code iteratively solves for a planet's density, pressure, gravity and adiabatic temperature profiles that are consistent with the pressures derived from hydrostatic equilibrium, the mass within a sphere, and the equation of state (EOS) of constituent minerals. We adopt models of $\sim$1000 shells in a spherically symmetric planet. We partition each modeled planet into a metal core composed of pure liquid-Fe, a rocky mantle composed of ${\rm MgO}$ and ${\rm SiO}_2$ compounds, and an outer water/ice layer. ExoPlex determines the mineralogy and density as determined by the EOS at each depth in the rocky mantle using the PerPlex Gibbs free energy minimizer package \cite{Conn09}. We adopt the thermally dependent EOS of liquid water of \cite{Stix90}, and for the ice layer, we have implemented an approximate equation of state including liquid water, ice I, ice VI and ice VII. 
\begin{centering}
\begin{figure}
    \centering
    \begin{tabular}{c}
    \includegraphics[width=7.5cm]{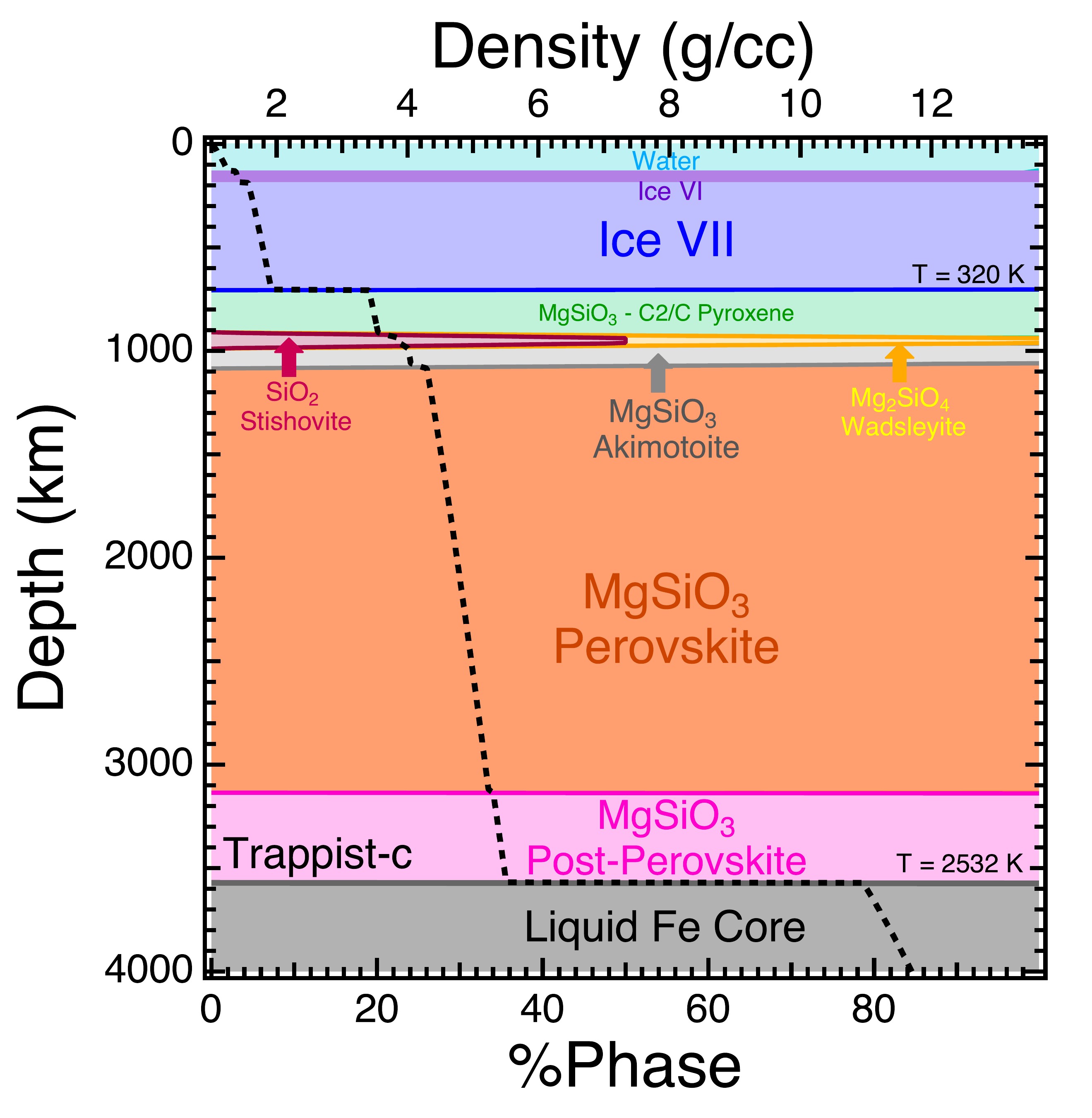}\\
    \includegraphics[width=7.5cm]{trap_f.pdf}
    \end{tabular}
    \caption{Phase diagrams with depth for the modeled best-fit interiors of \trap{}c and f with Fe/Mg = 0.8, Mg/Si = 1 and 8 and 50 wt\% \wat{}, respectively. Density as a function of depth is shown as a dashed line. Temperatures at the bottom each major boundary layer are included for reference. Note that due to the increased pressure at the water-mantle boundary ($\sim$25 GPa), no upper mantle phases are present in \trap{}g while due to the lower mass (and thus central pressure), no post-perovskite is present.
    }
    \label{fig:phases}
\end{figure}
\end{centering}
The refractory element composition of a terrestrial exoplanet composition is, to first-order, roughly that of its host star \cite{Bond10,Dorn15,Thia15,Unte16,Unte17}. The chemical composition of \trap{} is unknown, outside of its iron content: [Fe/H] = 0.04 $\pm$ 0.08 \cite{Gill16}. For these models then, we fixed the Mg/Si molar ratio to 1 and varied $0.5 < {\rm Fe}/{\rm Mg} < 1.3$ (note: the bulk Earth value is Fe/Mg = 0.9)\cite{McD03}. By varying Fe/Mg, we are effectively changing the core-to-mantle ratio of our modeled planet, whereas Mg/Si varies only the relative molar proportions of SiO$_2$, MgO and MgSiO$_3$ in the bulk of the mantle \cite{Unte16}. While a planet's Ca/Mg and Al/Mg ratios may play an important role in melting processes, mantle viscoelastic properties and phase equilibria, the mass-radius relation of a planet is relatively insensitive to their exact values \cite{Dorn15,Unte16}. We find that the Ca/Mg and Al/Mg ratios typically found in main sequence stars \cite{Hink14} change our calculated mass-radius relationships by less than a percent and thus we neglect these elements in our model for simplicity. 

While we neglect light elements in the core, the change to the mass-radius relation is only a few percent at Earth-like light element mass fractions ($\geq 10\%$\cite{Unte16}). Both water and light elements being present in the core lower the bulk density of a planet, and as such our proposed \trap{} water fractions should be considered upper limits. The inclusion of core light elements into our model would cause us to infer the \trap{} planets are drier than we report here, but our overall result of an increasing \textit{gradient} of water content with orbital distance is robust. Furthermore, this gradient is still present in our models when the uncertainties in the measured planet radius are included. 
ExoPlex follows the same iterative procedure seen in other papers \cite{Zapo69,Seag07,Unte16}. Given shells of fixed mass $d$ at various radii, $r$, ExoPlex calculates the gravity field, $g(r)$, and pressure gradient, $P(r)$, by integrating equation of hydrostatic equilibrium:
\begin{equation}
\frac{dP}{dr} = -\rho(r) \, g(r)
\end{equation}
where $\rho(r)$ is the density of the layer. $\rho(r)$ is then calculated using the equation of state of the stable constituent minerals within the layer as determined by PerPlex \cite{Conn09}. The positions $r$ of the shells is then recalculated using the volume calculated from the new density at depth and shell mass. This process is then iterated until convergence, which we define as the change in density in every shell between iterations does not change by one part in 10$^{-6}$. Because the rock relatively incompressible, this procedure converges within 20 iterations. ExoPlex is able to self-consistently calculate phase equilibria across a wide range of P-T space for a given change in bulk composition, rather than ad-hoc dictating the mineralogy as in previous studies (e.g. \cite{Zeng13, Unte16}). Thus it is more akin to the models developed in \cite{Dorn15}. We adopt the thermally dependent EOS formalisms of \cite{Stix05} for the mantle, \cite{Ander94} for the liquid Fe-core and The International Association for the Properties of Water and Steam (IAPWS) formalism for determining water/ice phase equilibria. We self-consistently calculate adiabatic geotherms starting with a potential temperature of 300 K for the water layer, mantle potential temperature of 1700 K, and core temperature of 2000 K greater than the mantle-core boundary temperature. ExoPlex output calculations for the mineralogy, density and temperature of \trap{}c and f with 8 and 50 wt\% water, respectively, with Fe/Mg = 0.8 and Si/Mg = 1 are shown in Supplementary Figure 1. Due to the lack of Ca and Al in our model, no garnet phases are present, which leads to the formation of stishovite and wadsleyite in the upper-mantles of \trap{}c despite Mg/Si = 1. 

\subsection{Hypatia Catalog}
\begin{centering}
\begin{figure}
    \centering
    \includegraphics[width=.75\linewidth]{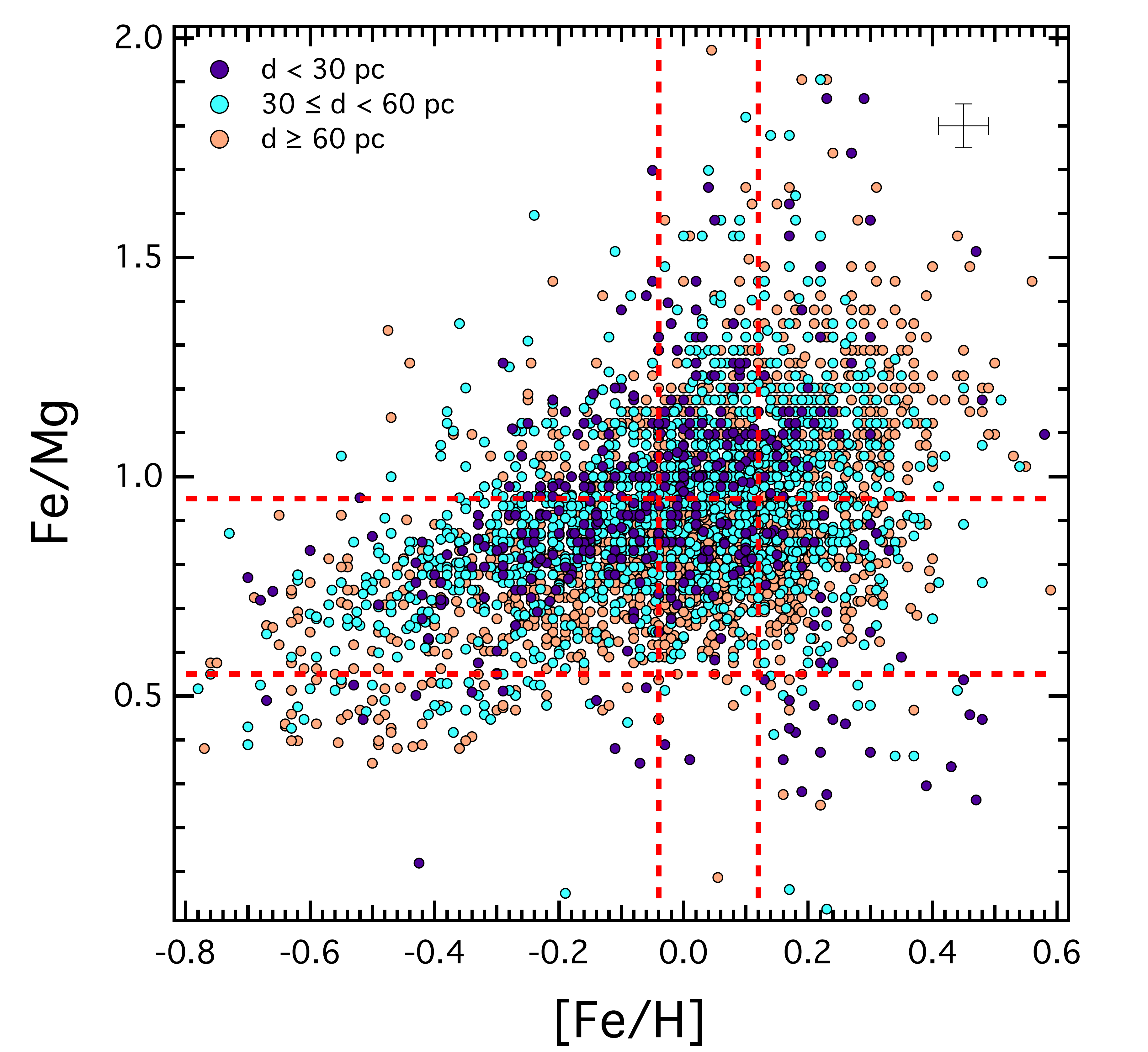}
    \caption{Range of Fe/Mg with respect to Solar normalized [Fe/H] for the full Hypatia Catalog sample. The dashed lines show the stellar compositions within 1$\sigma$ of the reported metallicity [2] and the Fe/Mg used in this analysis. Each star is color-coded to show distance from the Sun and the respective error (namely, the average of the error as reported by each Hypatia data set that measured that element [5]) is given in the top right corner of the plot.
}
    \label{fig:stars}
\end{figure}
\end{centering}

\begin{centering}
\begin{figure}
    \centering
    \includegraphics[width=.75\linewidth]{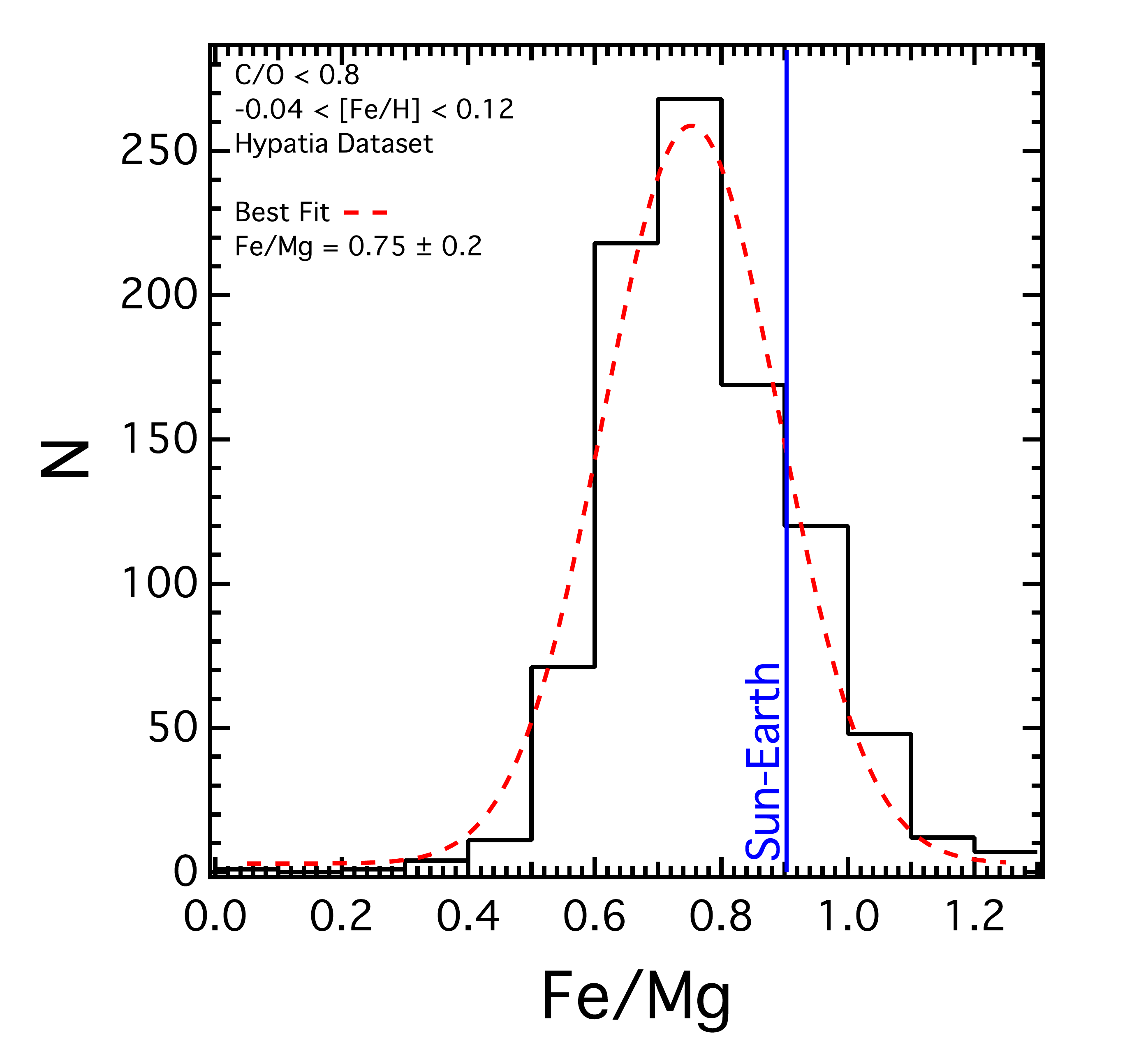}
    \caption{Histogram of Fe/Mg for a subsample of the Hypatia catalog with -0.04 $\leq$ [Fe/H] $\leq$ 0.12 [2] and C/O $\leq$ 0.8. The best fit gaussian distribution of Fe/Mg = 0.75 $\pm$ 0.2 is shown as a red dashed line. The Sun and Earth's Fe/Mg are shown as a blue line (Fe/Mg $\approx$ 0.9) [42,34]. }
    \label{fig:histogram}
\end{figure}
\end{centering}
The Hypatia Catalog \cite{Hink14,Hink16,Hink17} is an amalgamate dataset composed of stellar abundance measurements from $\sim$200 literature sources. Hypatia currently has $\sim$65 elements as found in $\sim$6000 FGK-type stars all found within 150 pc of the Sun. In order to put all of the abundance determinations on the baseline, all of the measurements were renormalized to the same solar abundance scale, namely \cite{Ande89} where log $\epsilon$(Fe) = 7.50. This solar normalization is the same that was used for \trap{} by \cite{Gill16}. The standard range of Fe/Mg found within Hypatia for well agreed upon, thin disk stars \cite{Hink16} are shown in Supplementary Figure 2, where the dashed lines show the compositions used within this analysis. Supplementary Figure 3 shows the Fe/Mg distribution of stars within 1$\sigma$ of the measured metallicity of \trap{}\cite{Gill17}. We therefore adopt this range of compositions for our interior models: 0.5$leq$Fe/Mg $leq$1.3. The Sun and Earth's Fe/Mg is approximately 0.9 \cite{Ande89,McD03}. The Hypatia Catalog can be found online at www.hypatiacatalog.com.

\subsection{Snow lines in M dwarf disks}
The position of the snow line in a protoplanetary disk depends on how the disk intercepts and reradiates stellar radiation, which means it depends on disk structure as well as stellar luminosity. Because of the disk's shielding ability, temperatures at a given radius in a protoplanetary disk are generally much lower than the temperatures of blackbodies in free space at the same radius. However, especially for M stars, stellar luminosity is much greater during the disk lifetime than several Gyr later. To compute where the snow line should be, we assume a passively heated (non-accreting) disk and calculate the temperature where the temperature is approximately that of the sublimation temperature of water ice. For the solar nebula, this temperature is, $T = 170 \, {\rm K}$. Given the larger surface density in an M-dwarf disk, however, this temperature will be higher than in the solar nebula. Using the equations found in \cite{Maue03}, we find that for a surface density of $1.2\times10^{5}$ g cm$^{-2}$, the vapor pressure of (half the) water at 0.24 AU is about 1 Pa, and the sublimation temperature is $\sim$212 K.  Following \cite{Kenn08}, we use the formulation of CG97\cite{Chia97}, but we do not assume parameters relevant to G stars. CG97 assume a disk in hydrostatic equilibrium whose radiating surface lies at height
\begin{equation}
H = 4 \, \frac{C}{\Omega} = 4 \, \left( \frac{k T}{\bar{m}} \right)^{1/2} \,
 \left( \frac{G M_{\star}}{r^3} \right)^{-1/2} 
\label{eq:height}
\end{equation}
above the midplane, $C$ being the sound speed and $\Omega$ the Keplerian orbital frequency at a
distance $r$ from the star of mass $M_{\star}$. Other variables have their usual meanings.
Starlight impinges on the surface at a glancing angle $\alpha$, where:
\begin{equation}
\alpha = \frac{d H}{dr} - \frac{H}{r} = r \, \frac{d}{dr} \left( \frac{H}{r} \right).
\end{equation}
The disk temperature at radius $r$ is easily found by balancing absorption of starlight against
disk emission, to find
\begin{equation}
T(r) = T_{\star} \, \alpha^{1/4} \, \left( \frac{r}{R_{\star}} \right)^{-1/2},
\label{eq:temp}
\end{equation}
where the star has temperature $T_{\star}$ and radius $R_{\star}$, and therefore luminosity
$L_{\star} = 4\pi R_{\star}^2 \, \sigma T_{\star}^4$.
Note that Equation 1 of CG97 incorrectly divides $\alpha$ by a factor of 2, reducing the temperature by about 20\%.
By combining these equations, it is straightforward to show that
\begin{equation}
T(r) = T_{\star} \,
 \left[ \frac{8}{7} \, \left( \frac{k T_{\star} R_{\star}}{G M_{\star} \bar{m}} \right)^{1/2} \right]^{2/7} \,
 \left( \frac{ r }{1 \, {\rm AU}} \right)^{-3/7}.
\end{equation}
For \trap{} we assume $M_{\star} = 0.08 \, M_{\odot}$ and therefore
$T_{\star} = 3000 \, {\rm K}$ throughout its descent onto the main sequence \cite{Bara02}.
At 10 Myr, its luminosity is $L_{\star} \approx 0.01 \, L_{\odot}$, corresponding to a radius $R_{\star} = 0.37 \, R_{\odot}$, yielding
\begin{equation}
T(r) = 62.6 \, \left( \frac{ L_{\star} }{ 0.01 \, L_{\odot} } \right)^{2/7} \,
               \left( \frac{ M_{\star} }{ 0.08 \, M_{\odot} } \right)^{-1/7} \,
               \left( \frac{ r }{ 1 \, {\rm AU} } \right)^{-3/7} \, {\rm K}.
\label{eq:temptwo}
\end{equation}
The location where $T = T_{\rm snow} = 170 \, {\rm K}$ for the solar nebula, however due to the higher surface density is therefore
\begin{equation}
r_{\rm snow} \approx 0.06 \, \left( \frac{ M_{\star} }{ 0.08 \, M_{\odot} } \right)^{-1/3} \,
                        \left( \frac{ L_{\star} }{ 0.01 \, L_{\odot} } \right)^{2/3} \, {\rm AU}.
\label{eq:rsnow2}
\end{equation}
Note that the stellar temperature and radius enter only through the stellar luminosity, so this
formula is generally applicable, and would predict $r_{\rm snow} \approx 0.14 \, {\rm AU}$ in the Kepler-32 system at 10 Myr, and $r_{\rm snow} \approx 0.9 \, {\rm AU}$ around the Sun at about 2 Myr, when $L \approx 1 \, L_{\odot}$ and $T_{snow} = 170 K$\cite{Bara02}. For the \trap{} system, it is clear that even at $t = 10 \, {\rm Myr}$ the snow line was well outside the current positions of the planets. It is important to note too that the inclusion of accretional heating in the disk would move the snow line even further out, requiring more migration than proposed here. If disk accretion were a significant source of heating, then the snow line would lie even farther from the star, strengthening the conclusion that the \trap{} planets are currently well inside the snow line of the disk they formed from, and almost certainly migrated inward.

\subsection{Migration in M Dwarf disks}
\cite{Swift13} argued for migration in the Kepler-32 system, an M dwarf star with 5 known transiting planets. The Kepler-32 planets far exceed the isolation masses of any reasonable disk at their present locations, are in coupled mean-motion resonances and also have significant volatile content despite being well inside the predicted locations of the water snow line (and in some cases even the silicate sublimation radius). Likewise, the \trap{} planets are in a remarkable resonant chain, with the period ratios all within 1\%, but all slightly larger than, integer ratios \cite{Gill17,Wang17}. They have large combined mass, $\sim4 \, M_{\oplus}$, mostly of rock. Applying a rock-to-gas ratio 0.005 \cite{Lodd03} yields a mass of gas $\sim0.0025 \, M_{\odot}$, about 3\% of the mass of \trap{} itself. This implies that the planets comprise most of the mass of the protoplanetary disk, not just the portion inside 0.06 AU. In fact, the implied surface density of gas in the disk between 0.013 AU and 0.018 AU (the annulus associated with \trap{}c) is $\sim 1.8 \times 10^7 \, {\rm g} \, {\rm cm}^{-2}$, which if the disk temperature is $T \approx 200$ K at 0.015 AU, suggests a Toomre parameter $Q \approx 0.7$. This Toomre parameter is below the threshold of $Q \approx 0.9$ for gravitational instability in M dwarf disks \cite{Back16}. As such, if the \trap{} planets formed in situ as in \cite{Quar17}, rather than migrated after formation, gas-rich Jupiter-like planets would be expected. Hence, the formation of the \trap{} in place is unlikely as this scenario would produce a gravitationally unstable disk and inconsistent with the observed masses and compositions of the planets themselves. Dynamical rearrangement in these systems may be possible, however given the circular orbits, very low mutual inclinations and period ratios slightly larger than 3:2 are highly suggestive that the \trap{} system did not undergo any such rearrangement. As our water mass fraction results are dependent on only mass and radius measurements, and thus dynamical history does not change our inferred compositions of the \trap{} system.

The discovery of multi-planet systems allows for the construction of minimum-mass solar nebula models for protoplanetary disks \cite{Kuch04, Gaid17}. Similar to those constructed for the solar nebula \cite{Desc07}, one must account for migration and consider the starting locations of the planets to estimate the disk surface density. If the \trap{} planets formed within $\sim$ 10 Myr, each of the planets' orbits must have decreased a factor of $\sim$4. Then, we may estimate the surface density in the disk as: $\Sigma \sim (4 \, M_{\oplus}) (0.005)^{-1}  /  [ \pi (4 * 0.06 \, {\rm AU})^2 ]$, or $\Sigma \sim 1.2 \times 10^5 \, {\rm g} \, {\rm cm}^{-2}$ inside  $\sim$0.24 AU. This surface density is orders of magnitude lower than the surface density estimated assuming \trap{}c formed in place, but it is still much higher than the surface densities inferred by similar reasoning for the solar nebula ($< 10^4 \, {\rm g} \, {\rm cm}^{-2}$). Even if the \trap{} planets formed within 3 Myr, so that the snow line at the time of their formation was $\sim8$ times farther from the star than \trap{}d's current orbit, the inferred surface densities still exceed those in the solar nebula.

It is encouraging and interesting that a similar conclusion was recently reached by \cite{Gaid17}, who constructed an average ``minimum-mass solar nebula'' estimate of $\Sigma$ in the protoplanetary disks that gave rise to {\it Kepler}-observed systems around M dwarfs. The models of \cite{Gaid17} include cases in which inward migration of planets is assumed, but with ad hoc assumptions about the starting locations of the planets. By identifying the break between less-icy and more-icy planets and comparing to the location of the snow line, we avoid relying on assumptions about the initial orbits of the planets. Nevertheless, the surface densities we infer for \trap{} are nearly identical to those they derived assuming migration. Despite its lower mass, \trap{}'s disk appears to have had surface densities comparable to those in other M dwarf disks, and higher than the solar nebula. We expect planet formation in M dwarf disks to proceed rapidly, perhaps forming planetary embryos directly as proposed by \cite{Chia13} and \cite{Gaid17}.

Surface density also plays a role in migration rate. Type I migration of $\sim1 \, M_{\oplus}$ planets, like those in the \trap{} system, leads to a change in semi-major axis at a rate $\dot{a} \approx -5 \, r \, \Omega \, (M_{\rm p} / M_*) \, ( \Sigma \,  r^2 / M_*) \, (h / r)^{-2}$, where $h$ is the scale-height of the disk, $M_{\rm p}$ and $r$ the mass and location of the planet, and the orbital frequency, $\Omega$ \cite{Tana02}. The factor $(\Sigma r^2 / M_{\star})$ appears to have been an order of magnitude greater in the \trap{} disk than in the solar nebula, and the factor $(h/r)^{-2}$, scaling with disk temperature, also is greater. We expect type I orbital migration to operate more than an order of magnitude faster in M dwarf disks. Type I migration may have played a limited role in the formation of the Solar System (as in the Grand Tack model\cite{Wals11}), but it seems to have played a significant role in the disks of \trap{}, Kepler-32, and potentially other M dwarf systems.

\end{document}